\documentclass[conference]{IEEEtran}
\usepackage{graphicx,bm,amsmath,amsfonts}

\newcommand{\lb}{\left}
\newcommand{\rb}{\right}

\newcommand{\Real}{\mathbb{R}}
\newcommand{\Natural}{\mathbb{N}}

\newcommand{\Tr}{\mathop{\mathrm{Tr}}}
\newcommand{\extr}{\mathop{\mathrm{Extr}}}

\newcommand{\Rr}{\bm{R}_{\mathrm{r}}}
\newcommand{\Rt}{\bm{R}_{\mathrm{t}}}

\usepackage{cite}

\begin{document}

\title{Statistical Mechanical Analysis of Compressed Sensing Utilizing 
Correlated Compression Matrix}

\author{\IEEEauthorblockN{Koujin Takeda}
\IEEEauthorblockA{Department of Computational Intelligence and\\ 
Systems Science\\ 
Tokyo Institute of Technology\\ 
Yokohama 226-8502, Japan\\
E-mail: takeda@sp.dis.titech.ac.jp
}
\and
\IEEEauthorblockN{Yoshiyuki Kabashima}
\IEEEauthorblockA{Department of Computational Intelligence and\\ 
Systems Science\\ 
Tokyo Institute of Technology\\ 
Yokohama 226-8502, Japan\\
E-mail: kaba@dis.titech.ac.jp
}
}


\maketitle

\begin{abstract}
We investigate a reconstruction limit of compressed sensing
for a reconstruction scheme based on 
the $L_1$-norm minimization
 utilizing a correlated 
compression matrix with a statistical mechanics method.
We focus on the compression matrix modeled as the 
Kronecker-type random matrix studied in research on multiple-input multiple-output wireless 
communication systems. We found that strong one-dimensional correlations between 
expansion bases of original information slightly degrade reconstruction performance.
\end{abstract}

\IEEEpeerreviewmaketitle


\section{Introduction}
A novel approach of data compression, termed compressed sensing (CS), has 
recently been drawing great attention. The central assumption of CS is
the sparsity of original information, which seems plausible for many real world signals. For exploiting 
this property, much effort has been paid in both research directions of theory and 
application \cite{CRT2006,D2006-1,CT2006}.

The basic idea of CS is summarized in the following linear equation: 
\begin{equation} 
\bm y = \bm F \bm x^{0}. 
\end{equation} 
(Throughout this article vectors and matrices are denoted in bold letters). $\bm x^{0} 
\in \Real^{N}$ denotes $N$-dimensional coefficient vector of the 
original information $\bm f \in \Real^{N}$ expanded over the basis ${\bm \phi}_i \in 
\Real^{N}$ $(1 \le i \le N)$, namely $\bm f = \sum_{i} x^{0}_{i} {\bm \phi}_i $, 
and $\bm y\in \Real^{P}$ is a $P$-dimensional vector, which describes compressed 
information available from $P$ times observations with observation vector $\bm 
\psi_i$, $\bm y_i = \bm f \cdot {\bm \psi}_i$ $(1 \le i \le P)$. $\bm F$ is a 
$P$-by-$N$ compression matrix whose element is given by $F_{ij} = \bm \psi_i \cdot 
\bm \phi_j$. In this article the compression matrix $\bm F$ is regarded as random, 
which means that we are dealing with random expansion bases and random 
observation vectors. The compression rate is defined by $\alpha \equiv P/N< 1 $. The 
original coefficient $\bm x^0$ is sparse and modeled by the distribution, 
\begin{equation} 
\label{eq:P0dist} P(x^0_i) = (1 - \rho) \delta(x^0_i) + \rho \exp \left( - (x^{0}_i)^2 /2 \right) / 
\sqrt{2\pi}, 
\end{equation} 
that is, $\rho$ represents the density of non-zero coefficients.
 Under the above setting, 
 $L_1$-norm minimization offers an appropriate feasible algorithm for 
reconstruction of the original coefficient (termed $L_1$-norm reconstruction), 
\begin{equation} 
{\rm minimize} \parallel \!\! \bm x \!\! \parallel_1 \ \ {\rm subject \ to \ \ } \bm y(=\bm F 
{\bm x}^0)= \bm F \bm x, 
\end{equation} 
where $\parallel \!\! \bm x \!\! \parallel_p = \lim_{\epsilon \rightarrow + 0} 
\sum_{i} |x_i|^{p+\epsilon}$. The remaining problem is whether the solution $\bm 
x$ coincides with the original coefficient $\bm x_0$. We can expect that below a 
certain critical value of the compression rate $\alpha_c$, the original coefficient ${\bm 
x}^0$ cannot be reproduced even if we make use of the $L_1$-norm reconstruction. 
The aim of this article is to evaluate this critical value $\alpha_c$ in the limit of $P,N 
\rightarrow \infty$ (and $\alpha={\rm const}.$) utilizing a statistical mechanics 
method.

By the way, this problem is quite similar to the performance 
evaluation problem of linear vector channels in wireless communication,
 and accordingly the analysis scheme with the 
statistical mechanical approach for code-division multiple-access (CDMA) or 
multiple-input multiple-output (MIMO) communication \cite{T2002,TUK,HTK} can be 
applied in the limit $P,N \rightarrow \infty$. Kabashima
et al. have already investigated the 
performance of the $L_1$-norm reconstruction (to be precise general
$L_p$-norm, though $p \le 1$ is the reconstructable case) 
using a statistical mechanical method in a basic scenario in \cite{KWT},
where $\bm F$ is 
composed of independently and identically distributed (i.i.d.) random variables, and 
evaluated the reconstruction limit $\alpha_c$. The evaluation value accords 
with the one that has been assessed in \cite{D2006-2,DT2009} using combinatorial 
geometry methods. (As a related study, noisy compressed sensing was
investigated using replica method in \cite{RFG}, for which
perfect reconstruction is not possible as long as the noise intensity
is not negligible).

In this article, as a second step of the investigation, we consider a more advanced case in which 
$\bm F$ is provided as 
\begin{equation} 
\label{eq:defF} \bm F = \sqrt{\Rr} \bm \Xi \sqrt{\Rt}, 
\end{equation} 
and our goal is to evaluate critical value $\alpha_c$ for such $\bm F$.
Here, $\Rr$ and $\Rt$ are a $P$- and an $N$-dimensional square symmetric matrix, 
respectively. The square root of a square matrix $\bm A$ is defined as $\bm A = 
\sqrt{\bm A}^{T} \sqrt{\bm A}$. $\bm \Xi$ is a random $P$-by-$N$ rectangular 
matrix whose elements are i.i.d. Gaussian random variables of zero mean and variance 
$N^{-1}$. This random matrix $\bm \Xi$ effectively implies a situation in that the 
expansion bases and the observation vectors are statistically uncorrelated. In this 
modeling, the matrices $\Rt$ and $\Rr$ represent the correlations among the expansion 
bases $\bm \phi$ and those among the observation vectors $\bm \psi$, respectively. 
Random matrix of this type, $\bm F$, is known as the channel matrix
in the Kronecker model of the MIMO 
communication system, whose performance is investigated by Hatabu et
al. \cite{HTK} with a statistical mechanical scheme. Accordingly, by application of 
this method it is expected that the reconstruction limit of the 
$L_p$-norm reconstruction can also be estimated. In the subsequent sections we explain 
the details of the analysis.


\section{Replica analysis} 
In this section we describe the outline of the analysis.
 As we mentioned, the analysis is 
based on that for the Kronecker channel in the MIMO 
communication system \cite{HTK}, and the details of the analysis are also  
discussed in this work.

Following the discussions in \cite{KWT}, let us first define the cost function of the 
$L_p$-norm reconstruction using the quenched average of free energy, which is a 
standard technique for dealing with a random system in statistical mechanics, 
\begin{eqnarray} 
\label{eq:freeenergy} C_p & \equiv & - \lim_{\beta \rightarrow \infty} \lim_{n \rightarrow 0} 
\frac{\partial}{\partial n} \lim_{N \rightarrow \infty} \frac{1}{\beta N} \ln [ Z^n (\beta,\bm 
y)]_{\bm F, \bm x^0}, 
\end{eqnarray} 
where $[ \cdot ]_{\bm F, \bm x^0}$ denotes the average over the random matrix $\bm 
F$ and the original coefficient $\bm x^0$ with the distribution (\ref{eq:P0dist}). We 
also define the replicated partition function $Z^{n}(\beta,\bm y)$ for $n \in 
\Natural$ as 
\begin{eqnarray} 
\label{eq:ffff1} Z^{n}(\beta, \bm y) & \equiv & \prod_{a=1}^{n} \int {d \bm x^a} \exp (- \beta \parallel 
\!\! \bm x^a \!\! \parallel _p) \delta (\bm F (\bm x^a - \bm x^0)) \nonumber \\
&& \hspace{-2.0cm} = \prod_{a=1}^{n} \int {d \bm x^a} \lim_{\tau \rightarrow +0} \frac{1}{(\sqrt{2 \pi 
\tau})^{nP}} \nonumber \\
&& \hspace{-2.5cm} \times \exp \left[ - \sum_{a=1}^{p} \beta \parallel \!\! \bm x^a \!\! 
\parallel _p -\frac{1}{2 \tau} \sum_{a=1}^{n} (\bm x^a - \bm x^0)^{T} \bm F^{T} \bm F (\bm 
x^a - \bm x^0) \right]. \nonumber \\
\end{eqnarray} 
From these expressions, we easily see that the cost function $C_p$ is nothing but the 
minimized norm with the constraint $\bm y \equiv \bm F \bm x = \bm F \bm 
x^0$ for given $\bm y$. After performing the average over
 the random matrix $\bm \Xi$ in $\bm F$ 
we have 
\begin{eqnarray} 
\label{eq:term21}
 && \hspace{-0.5cm} \int d \bm F \int d \bm x^0 \prod_{a=1}^{n} \int {d \bm x^a} \lim_{\tau 
\rightarrow +0} \frac{1}{(\sqrt{2 \pi \tau})^{nP}} \nonumber \\
&& \hspace{-1cm} \times \exp \left[ -\frac{1}{2 \tau} \sum_{a=1}^{n} (\bm x^a - \bm x^0)^{T} 
\bm F^{T} \bm F (\bm x^a - \bm x^0) - \sum_{a=1}^{n} \beta \parallel \!\! \bm x^a \!\! 
\parallel _p \right] \nonumber \\
&=& \int d \bm x^0 \lim_{\tau \rightarrow +0} \frac{1}{(\sqrt{ 2 \pi \tau})^{nP}} \nonumber \\
&&\hspace{-0.5cm} \times \int d\bm Q \exp \lb[ N \Tr G_{\bm\Xi^{T}\Rr\bm\Xi} \lb(- 
\frac{1}{\tau} \bm S \rb) + \ln \Pi^{(n)} (\bm Q) \rb], 
\end{eqnarray} 
where 
$(\bm S)_{ab} \equiv Q_{ab} - Q_{a0} - Q_{0b} + Q_{00}$ 
and $\bm Q$ is an $n$-dimensional matrix defined by the constraint, 
\begin{eqnarray} 
\label{eq:defPi} \Pi^{(n)}(\bm Q) &\equiv& \prod_{a=1}^{n} \int d \bm x^{a} 
\lb\{\prod_{a=1}^{n} \delta(\bm x^{a T} \Rt \bm x^{a}-NQ_{aa})\rb\}
\nonumber \\&&\hspace{-2.3cm} \times \lb\{\prod_{a<b}^{n}\delta(\bm x^{a T} \Rt \bm x^{b}-NQ_{ab})\rb\} 
\lb\{\prod_{a=1}^{n} \delta(\bm x^{a T} \Rt \bm x^{0}-NQ_{a0})\rb\} \nonumber 
\\
&& \hspace{-2.3cm} \times \lb\{\delta(\bm x^{0 T} \Rt \bm x^{0}-NQ_{00})\rb\} \exp \left( - 
\sum_{a=1}^{n} \beta \parallel \!\! \bm x^a \!\! \parallel _p \right). 
\end{eqnarray} 
The function $G_{\bm \Xi^{T} \Rr \bm \Xi}$ is defined as 
\begin{eqnarray} 
\label{eq:Gfunc} G_{\bm \Xi^{T} \Rr \bm \Xi}(\bm A) &\equiv& - \frac{\alpha}{2}\int 
d\lambda \rho_{\Rr}(\lambda) \ln \lb(\bm I - \frac{\lambda}{\alpha} \bm A \rb). 
\end{eqnarray} 
The function $\rho_{\Rr}(\lambda)$ in the definition of $G_{\bm \Xi^{T} \Rr \bm 
\Xi}$ is the eigenvalue distribution of the matrix $\Rr$.

Assuming replica symmetry, let $q=Q_{ab}\ (a \ne b)$, $Q = Q_{aa}$, 
$m=Q_{a0}$, and $u=Q_{00}$. Here, $u$ is defined as $u \equiv N^{-1} \int 
\prod_{i} d x_i^{0} P(x^0) {\bm x^{0 T}} \Rt {\bm x}^0 = \frac{\rho}{N} \Tr \Rt. 
$ From these assumptions it follows that $S_{aa} = Q-2m+u$ and $S_{ab}=q-2m+u 
\ (a \ne b)$. By diagonalization of the matrix $\bm S$, we can evaluate the $G_{\bm 
\Xi^{T} \Rr \bm \Xi}$ -dependent part, 
\begin{eqnarray} 
&& \exp \lb[ N \Tr G_{\bm\Xi^{T}\Rr\bm\Xi} \lb(- \frac{1}{\tau} \bm S \rb) \rb] \nonumber 
\\
&=& \exp \lb[ N \left\{ G_{\bm\Xi^{T}\Rr\bm\Xi} \lb( - \frac{Q - q}{\tau} \rb) \rb. \rb. 
\nonumber \\
&& \hspace{-0.5cm} \lb. \lb. - \frac{ n ( q -2m + u)}{\tau} G'_{\bm\Xi^{T}\Rr\bm\Xi} \lb( - \frac{Q - q}{\tau} 
\rb) +O (n^2) \right. \rb. \nonumber \\
&& \lb. \left. + (n-1) G_{\bm\Xi^{T}\Rr\bm\Xi} \lb( - \frac{Q - q}{\tau} \rb) \right\} \rb]. 
\end{eqnarray} 
From the saddle-point method, the $\bm x$-dependent part, including the 
$L_p$-norm and the constraint, is expressed as,
\begin{eqnarray} 
\label{eq:Sn} &&\Pi^{(n)}(\bm Q) \nonumber \\
&\equiv& \extr_{\tilde{Q},\tilde{q},\tilde{m}} \left( \exp \lb\{ -NnQ\tilde{Q} - 
N\frac{n(n-1)}{2}q \tilde{q} -Nnm\tilde{m} \rb\} \right. \nonumber \\
&& \hspace{-0.3cm} \times \left. \int D \tilde{\bm z} \left( \int d \bm x \exp \lb[ \left( \tilde{Q} - 
\frac{\tilde{q}}{2} \right) \bm x^{T} \Rt \bm x \right. \right. \right. \nonumber \\
&& \left. \left. \left. \hspace{-3mm} + \bm x^{T} \sqrt{\Rt}^{T} \left( \tilde{m} \sqrt{\Rt} \bm 
x^{0} + \sqrt{\tilde{q}} \tilde{\bm z} \right) - \beta \parallel \!\! \bm x \!\! \parallel _p \rb] 
\right)^{n} \right), 
\end{eqnarray} 
where the interaction between replicas is removed by introducing auxiliary variable 
$\bm z$ (Hubbard-Stratonovich transformation). For simpler expression of
$C_p$, let us define 
new variables $\widehat{m} \equiv \beta^{-1} \tilde{m},
 \widehat{\chi} \equiv \beta^{-2} 
\tilde{q}, \chi \equiv \beta (Q - q),
\widehat{Q} \equiv \beta^{-1} (-2\tilde{Q} + 
\tilde{q})$ and the function 
\begin{equation} 
\phi_p(\bm h,\widehat{Q}) \hspace{-0.1cm} \equiv \hspace{-0.1cm} \frac{1}{N} 
\lim_{\epsilon \rightarrow +0} {\rm min}_{\bm x} \left\{ \frac{\widehat{Q}}{2} \bm x^{T} \Rt 
\bm x - \bm h^{T} \sqrt{\Rt} \bm x + \parallel \!\! \bm x\!\! \parallel _{p+\epsilon} \right\}. 
\end{equation} 
For $\beta \rightarrow \infty$, the $\bm x^0$-dependent part is rewritten as 
\begin{equation} 
\prod_{i} \int P(x_i^0) d x_i^0 \int D \tilde{\bm z} \exp \left( -\beta N n \phi_{p}( \widehat{m} 
\sqrt{\Rt} {\bm x}^{0} + \sqrt{\widehat{\chi}} \tilde{\bm z}, \widehat{Q}) \right). 
\end{equation} 
Combining these results and inserting the expression of the function $G_{\bm \Xi^{T} 
\Rr \bm \Xi}$ and its derivative in the limit $\tau \rightarrow +0$, we have the final 
expression of the cost function 
\begin{eqnarray} 
\label{eq:Cpfinal} C_p &=& \hspace{-0.5cm} 
\extr_{q,m,\chi,\widehat{Q},\widehat{m},\widehat{\chi}} \left( \left. \frac{ \alpha ( q -2m + 
u)}{2 \chi} + \left( \frac{\chi \widehat{\chi}}{2} - \frac{q \widehat{Q}}{2} + m \widehat{m} 
\right) \right. \right. \nonumber \\
&& \hspace{-1cm} + \left. \left\{ \prod_{i} \int d x_i^0 P(x_i^0) \int D \tilde{\bm z} 
\phi_{p}( \widehat{m} \sqrt{\Rt} {\bm x}^{0} + \sqrt{\widehat{\chi}} \tilde{\bm z}, 
\widehat{Q}) \right\} \right). \nonumber \\
\end{eqnarray} 
The cost function depends only on the correlation matrix $\Rt$ between expansion 
bases of the original information, and does not depend on the matrix $\Rr$ between 
observation vectors, which indicates that the observation procedure
is not essential for CS. This seems reasonable 
because sparsity of the original coefficient is significant and
observation is not for CS. 
Accordingly, we must concentrate only on the effect of the correlation matrix 
$\Rt$ on the $L_p$-norm reconstruction.

As mentioned above, $C_p$ is nothing but the minimized $L_p$-norm, and the 
expression Eq. (\ref{eq:Cpfinal}) tells us that the minimized $L_p$-norm is given 
by the solution of the extremization problem. The remaining problem is whether the 
original coefficient is correctly reconstructed typically from the solution of the 
extremization problem. Remembering 
the fact that $\bm x$ is the result of the reconstruction and $\bm x^0$ is the original 
coefficient, from Eq.(\ref{eq:defPi}) $q=m(=u)$ must hold when the reconstruction is 
successful. Therefore, the scheme for finding the reconstruction limit $\alpha_c$ is as 
follows: vary the parameter $\alpha$ (compression rate) and $\rho$
(density of non-zero coefficients), then solve the extremization problem, and 
examine whether the solution satisfies $q=m(=u)$.

We have completed the replica analysis as above, and the cost function 
(\ref{eq:Cpfinal}) we obtained describes the information of the $L_p$-norm 
reconstruction for arbitrary correlation matrices $\Rr$ and $\Rt$. (As you see 
$\Rr$ will eventually become irrelevant). 
However, one problem remains: the cost function 
$C_p$ includes the function $\phi_p (\bm h, \widehat{Q})$, which is defined by the 
minimization problem with $N$ variables, whose expression is not so simple (For 
$L_1$-norm this problem can be solved numerically in principle because the 
minimization function is unimodal). Fortunately, for certain classes of $\Rt$, this 
minimization problem can be expressed relatively simply, which allows us to evaluate 
the reconstruction limit in a tractable manner. In the case without correlation $\Rt=\bm 
I$, we see that the minimization problem with $N$ variables is reduced to the problem 
with a single variable, and the cost function changes to
the one obtained in \cite{KWT}. In the following section,
 we will give another simple but nontrivial example, 
for which the minimization problem is numerically tractable.


\section{example: adjacent correlation}
Let us consider that the correlation matrix $\Rt$ has a tridiagonal form, as 
discussed in \cite{HTK}, in the context of the MIMO communication system, 
defined by 

\begin{equation} 
\Rt = \left( 
\begin{array}{cccccc} 
1 & r & 0 &  \ldots & r \\
r & 1 & r &  \ldots & 0\\
0 & r & 1 &  \ldots & 0\\
\vdots & \vdots & \vdots & \vdots & \vdots\\
r & 0 & 0 &  \ldots & 1 \\
\end{array} \right). 
\end{equation} 
This corresponds to the case that only adjacent matrix components (or adjacent 
expansion bases) have correlation. This matrix can be decomposed as $\Rt = 
\sqrt{\Rt}^{T} \sqrt{\Rt}$ (Cholesky decomposition with boundary term), where 
\begin{eqnarray} 
\sqrt{\Rt} &=& \left( 
\begin{array}{cccccc} 
l_+ & l_- & 0 &  \ldots & 0 \\
0 & l_+ & l_- &  \ldots & 0\\
0 & 0 & l_+ &  \ldots & 0\\
\vdots & \vdots & \vdots & \vdots & \vdots\\
l_- & 0 & 0 & \ldots & l_+ \\
\end{array} \right). 
\end{eqnarray} 
Here $l_{\pm} \equiv (\sqrt{1+2r} \pm \sqrt{1-2r})/2$. In what follows we focus 
on the $L_1$-norm minimization. In the present case $u= \Tr \Rt / N = \rho$ holds, 
and the cost function is rewritten as 
\begin{eqnarray} 
C_1 \hspace{-2mm} &=& \hspace{-6mm} 
\extr_{q,m,\chi,\widehat{Q},\widehat{m},\widehat{\chi}} \left( \left\{ \frac{ \alpha ( q -2m + 
\rho)}{2 \chi} + \left( \frac{\chi \widehat{\chi}}{2} - \frac{q \widehat{Q}}{2} + m \widehat{m} 
\right) \right. \right. \nonumber \\
&& \left. \left. + \prod_{i} \int P(x_i^0) d x_i^0 \int \prod_j D \tilde{z}_j \phi_{1}( \widehat{\bm 
h}, \widehat{Q}) \right\} \right), 
\end{eqnarray} 
where $\widehat{h}_{i} \equiv \widehat{m} ( l_+ x_i^0 + l_- x_{i+1}^0) + 
\sqrt{\widehat{\chi}} \tilde{z}_i $ and $x^0$ is defined periodically as $x_{N+1}^0 
= x_{1}^0$. (For simplicity we denote $\widehat{\bm h} \equiv \{ \widehat{h}_1, 
\cdots, \widehat{h}_N\}$). The function $\phi_1(\widehat{\bm h},\widehat{Q})$ in 
the cost function $C_1$ can be transformed as 
\begin{eqnarray} 
\phi_1(\widehat{\bm h} ,\widehat{Q})& = & \frac{1}{N} {\rm min}_{\bm x} 
\left\{ \frac{\widehat{Q}}{2} \bm x^{T} \Rt \bm x - \bm {\widehat{\bm h}}^{T} \sqrt{\Rt} \bm 
x + \parallel \!\! \bm x \!\! \parallel _{1} \right\} \nonumber \\
& \hspace{-2.5cm} = & \hspace{-1.5cm} \frac{1}{N} \left\{ \frac{\widehat{Q}}{2} \sum_{i} 
(x_i^*)^2 + \widehat{Q} r \sum_{i} x_i^* x_{i+1}^* - \widehat{m} \sum_{i} x_i^* x_i^0 \right. 
\nonumber \\
&& \hspace{-2cm} \left. - \widehat{m} r \sum_{i} x_i^* x_{i+1}^0 - \sqrt{\widehat{\chi}} 
\sum_{i} \tilde{z}_i (l_+ x_i^* + l_- x_{i+1}^*) + \sum_{i} | x_i^* | \right\}. \nonumber \\
\end{eqnarray} 
The variables $x,h$ are also defined periodically, $x_0 = x_N$, $x_{N+1}=x_1$, $h_0 
= h_N$, and $h_{N+1}=h_1$. $x_i^*$ is given by the solution of the minimization 
problem, namely for each $i$ 
\begin{eqnarray} 
\label{eq:extremizex} &&\hspace{-0.7cm} \frac{\partial}{\partial x_i^*} \phi_1(\bm h,\widehat{Q})= 
(\widehat{Q} x_i^* - \widehat{m} x_i^0 ) + r (\widehat{Q} x_{i-1}^* - \widehat{m} x_{i-1}^0) 
\nonumber \\
&&\hspace{-0.5cm} + r (\widehat{Q} x_{i+1}^* - \widehat{m} x_{i+1}^0) - \sqrt{\widehat{\chi}} (l_+ \tilde{z}_i + 
l_- \tilde{z}_{i-1} ) + {\rm sgn} (x_i^*) \nonumber \\
&& \hspace{-0.5cm}= 0 
\end{eqnarray} 
is satisfied. As seen above, the minimization problem for each $i$ includes only 
variables with three sequential indices $i-1, i$, and $i+1$, which indicates that the 
minimization problem is on a one-dimensional chain. It should be noted that the 
minimization function is unimodal, and sequential minimization for each variable 
enables us to find the minimum when we try to search it numerically. The 
computational cost of this procedure is $O(N)$ and feasible.

The extremization condition of the cost function can be expressed using the solution 
of the minimization problem, denoted by $x^*$, 
\begin{eqnarray} 
\label{eq:saddle} &&\hspace{-0.5cm} \widehat{Q} = \widehat{m} = \frac{\alpha}{\chi},  \
 \ \ \ 
\widehat{\chi} = \frac{\alpha(q - 2m + \rho)}{\chi^2}, \nonumber \\
&&\hspace{-0.5cm} q = \frac{1}{N} \prod_{i} \int D \tilde{z}_i d x_i^0 P(x_i^0) \left. \sum_{i} x_i^* (x_{i}^{*} + r 
x_{i-1}^{*} + r 	 x_{i+1}^{*}), \right. \nonumber \\
&&\hspace{-0.5cm} m = \frac{1}{N} \prod_{i} \int D \tilde{z}_i d x_i^0 P(x_i^0) \hspace{0cm} \sum_{j} x_j^* 
( x_j^0 + r x_{j-1}^0 + r x_{j+1}^0), \nonumber \\
&&\hspace{-0.5cm} \chi = \frac{1}{\sqrt{\widehat{\chi}}N} \prod_{i} \int D \tilde{z}_i d x_i^0 P(x_i^0), \left. 
\sum_{j} \ \tilde{z}_j ( l_+ x_j^* + l_- x_{j+1}^*). \right. \nonumber \\
\end{eqnarray} 
Remember that $x_i^*$ depends on 
$\widehat{Q},\widehat{m},\sqrt{\widehat{\chi}},\tilde{z}_i,x_i^0$ through Eq. 
(\ref{eq:extremizex}).

The next issue is the reconstruction limit as discussed in \cite{KWT}. As mentioned 
before, if the $L_1$-norm reconstruction works successfully, $q=m(=\rho)$ holds and 
the right hand side of the equation for $\widehat{\chi}$ in Eq. (\ref{eq:saddle}) 
vanishes. The extremization conditions for $\widehat{\chi},q,m$ can be combined by 
using a new variable $\widehat{x}_i \equiv x_i^*- x_i^0 $, 
\begin{equation} 
\widehat{\chi} = \frac{\alpha}{\chi^2 N} \prod_{i} \int D \tilde{z}_i d x_i^0 P(x_i^0) d x_i^0 
\sum_j \widehat{x}_j (\widehat{x}_j + r \widehat{x}_{j-1} + r \widehat{x}_{j+1}). 
\end{equation} 
From this expression it follows that $\widehat{x}_i=0$ (namely $x_{i}^* = x_{i}^0$) 
is obtained from the extremization conditions as a solution of successful reconstruction. 
On the other hand, by inserting $\widehat{m}=\widehat{Q}=\alpha/\chi$ into Eq. 
(\ref{eq:extremizex}), 
\begin{eqnarray} 
\label{eq:extremizex2} \hspace{-1cm} \frac{\partial}{\partial x_i^*} \phi_1(\bm h,\widehat{Q}) 
&=& \frac{\alpha}{\chi} \left( \widehat{x}_i + r \widehat{x}_{i-1} +r \widehat{x}_{i+1} \right) 
\nonumber \\
&& \hspace{-1cm} - \sqrt{\widehat{\chi}} (l_+ \tilde{z}_i + l_- \tilde{z}_{i-1} ) + {\rm sgn} 
(\widehat{x}_i + x_i^0 ) = 0. 
\end{eqnarray} 
This equation indicates that in the limit $\chi \rightarrow 0$ $\widehat{x}_i$ should 
vanish faster than $O(\chi)$ in order for the solution 
$\widehat{x}_i=0$ to exist. (In this case $\widehat{\chi}$ is $O(1)$). From 
the insight above, we rescale the variable as $\widehat{x} \rightarrow (\chi /\alpha) 
\widehat{x}$. The remaining equations in terms of 
$\chi,\widehat{\chi},\widehat{x}_i$ are 
\begin{eqnarray} 
\label{eq:finalresult} && \hspace{-0.7cm} \widehat{\chi} = \frac{1}{\alpha N} \prod_{i} \int D 
\tilde{z}_i d x_i^0 P(x_i^0) \sum_j \widehat{x}_j ( \widehat{x}_j + r \widehat{x}_{j-1} + r 
\widehat{x}_{j+1}), \nonumber \\
&& \hspace{-0.7cm} \chi = \chi \hspace{-0.1cm} \left( \frac{1}{\alpha
\sqrt{\widehat{\chi}} N} \prod_{i} \int D 
\tilde{z}_i d x_i^0 P(x_i^0) \sum_{j} \ \tilde{z}_j ( l_+ \widehat{x}_j + l_- \widehat{x}_{j+1}) 
\hspace{-0.1cm}\right)\hspace{-0.1cm}, \nonumber \\
&& \hspace{-0.7cm} \frac{\partial}{\partial x_i^*} \phi_1(\bm h,\widehat{Q}) = \left( \widehat{x}_i + r 
\widehat{x}_{i-1} +r \widehat{x}_{i+1} \right) \nonumber \\
&& \hspace{0.5cm} - \sqrt{\widehat{\chi}} (l_+ \tilde{z}_i + l_- \tilde{z}_{i-1} ) + {\rm sgn} 
\left( \frac{\chi}{\alpha} \widehat{x}_i + x^0_i \right) = 0. 
\end{eqnarray} 
The second equation has two solutions: $\chi=0$ and the factor in the 
bracket is unity. The first solution $\chi=0$ corresponds to successful 
reconstruction, $x_i=x_i^*$ (note that $\widehat{x}$ is rescaled), and the second 
(which satisfies $\chi \ne 0$) amounts to unsuccessful reconstruction. The equation for 
the threshold is obtained by inserting $\chi=0$ to the second solution. In 
conjunction with the remaining two equations, we finally have the equations for the 
reconstruction limit (Note that in the evaluation of the reconstruction limit, we must 
perform multiple integrals in the equations for $\chi$ and $\widehat{\chi}$, and we 
can use the Monte Carlo method in numerical evaluation).

Equation (\ref{eq:finalresult}) is the main result. By the procedure mentioned above, we can estimate the reconstructions $\alpha_c$ as 
the function of $\rho$ (density of non-zero coefficients) and 
$r$ (correlation parameter in the compression matrix). For $\Rt = \bm I$ (without 
correlation), we can recover the reconstruction limit obtained in \cite{KWT} directly 
from the expression of Eq. (\ref{eq:finalresult}).


\section{evaluation of reconstruction limit}
\begin{figure} 
\begin{picture}(110,140) \put(20,-10.5){\includegraphics[width=0.40\textwidth]{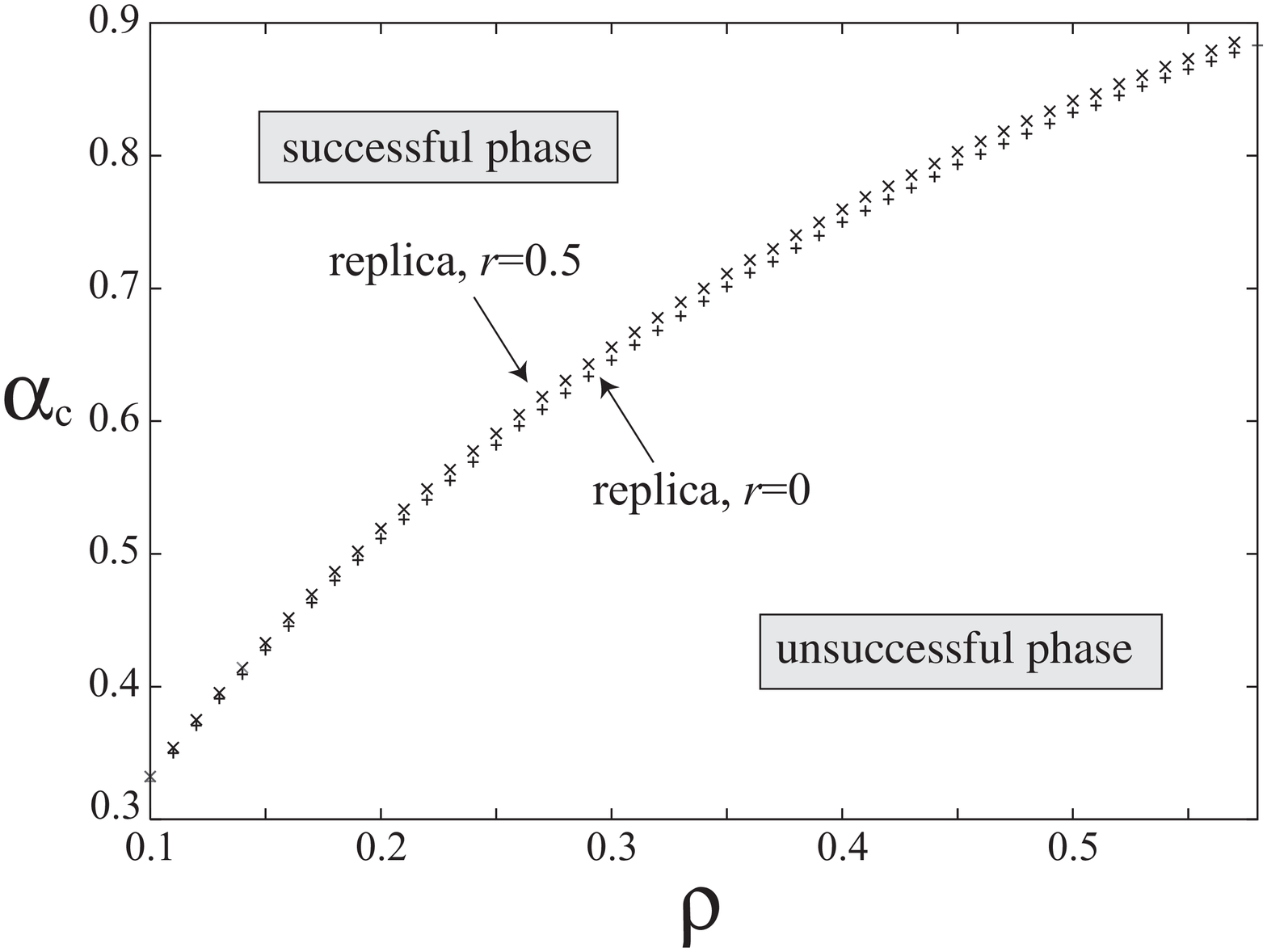}} 
\end{picture} 
\caption{Reconstruction limit $\alpha_c$ as a function of $\rho$.
We compare the cases $r=0.5$ $(\times)$ and $r=0$ ($+$, uncorrelated).
 We set the dimension of the original coefficient as $N=10^6$. For the evaluation of the multiple integrals in Eq. 
(\ref{eq:finalresult}), we perform the average over 100 samples of $\tilde{\bm z},\bm x^0$. The difference between two results is very small.} 
\label{fig:figure1} 
\end{figure}

\begin{figure} 
\begin{picture}(110,140) \put(10,-10.5){\includegraphics[width=0.40\textwidth]{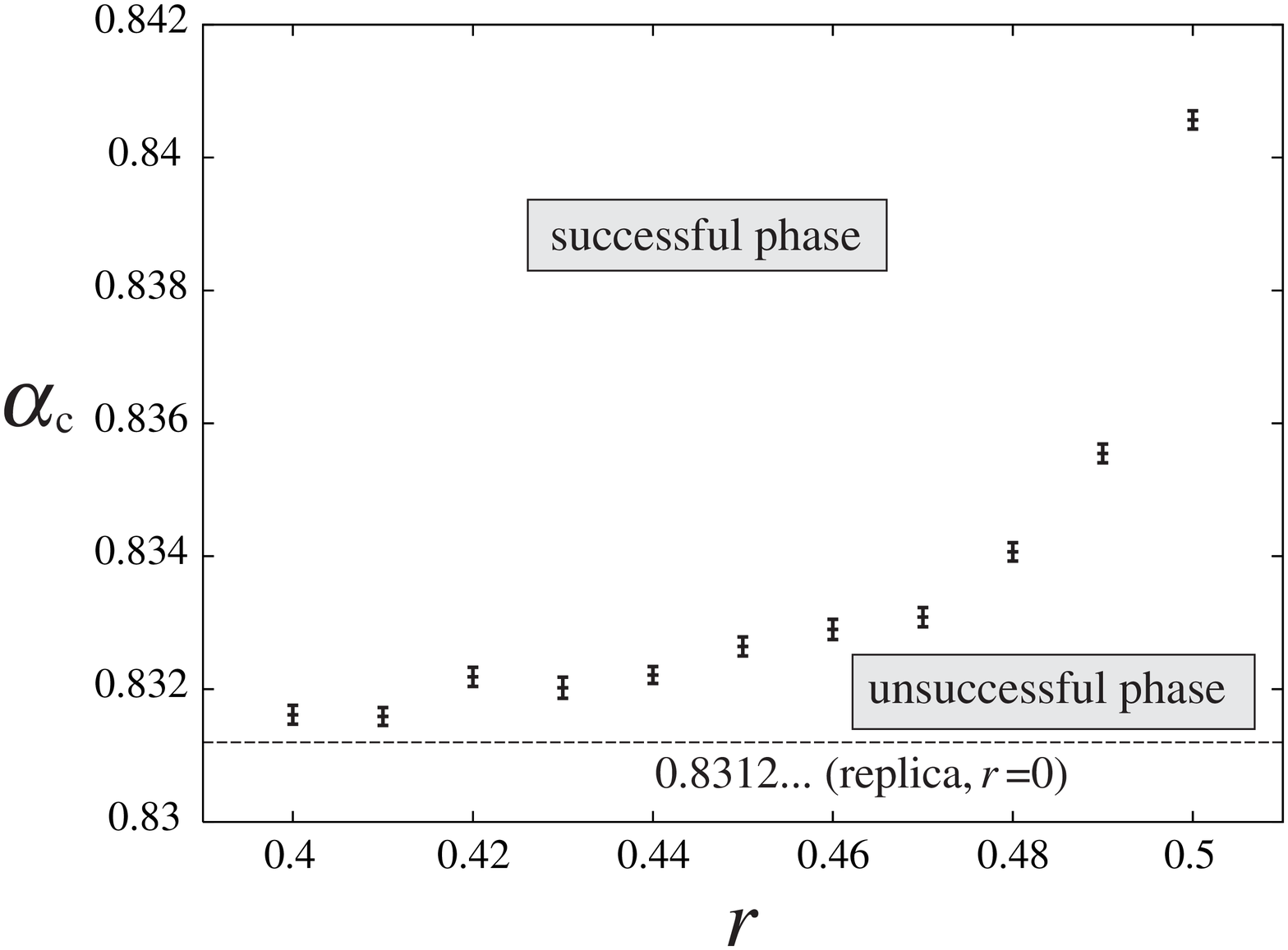}} 
\end{picture}
 \caption{Dependence of reconstruction limit $\alpha_c$ on correlation 
parameter $r$. We set $\rho=0.5$ and the dimension of the original 
coefficient as $N=10^6$. For the evaluation of the multiple integrals in Eq. 
(\ref{eq:finalresult}), we perform the average over 100 samples of $\tilde{\bm z},\bm 
x^0$. In the vicinity of $r=0.5$, the deviation of $\alpha_c$ from the value for $r=0$ is 
clearly observed. (For $\rho=0.5$, the reconstruction limit is estimated as 
$\alpha_c=0.8312...$ \cite{KWT}.) For $r<0.4$ the deviation is not visible and the 
result is not shown in this figure. At $r=0.5$, the criticality is estimated as 
$\alpha_c=0.84057(14).$} 
\label{fig:figure2} 
\end{figure}

For the adjacent correlation discussed above, we estimate the 
reconstruction limit by using the result of the replica analysis in Eq. 
(\ref{eq:finalresult}). In Fig. \ref{fig:figure1}, the dependences of
reconstruction limit $\alpha_c$ on the density $\rho$ are
shown for uncorrelated $(r=0)$ and correlated $(r=0.5)$ cases.
The difference between two results are very small over all region of
$\rho$, which indicates the effect of adjacent correlation is very small.
In Fig. \ref{fig:figure2} the dependence 
on the correlation parameter $r$ is depicted for $\rho=0.5$.
In the case without correlation, Kabashima et al. obtained 
$\alpha_c=0.8312...$ for $\rho=0.5$ \cite{KWT}. In the region of small $r$, we 
cannot observe the deviation of $\alpha_c$ from the uncorrelated case $r=0$. On the 
other hand, for the strongly-correlated case $r=0.5$, we observe a slight increase in 
$\alpha_c$, which implies that strong correlation worsens the performance of the 
$L_1$-norm reconstruction. For $r=0.5$, $\alpha_c$ is estimated as 
$\alpha_c=0.84057(14)$, indicating that the performance falls about $1\%$ from 
$r=0$ in terms of the reconstruction limit.

\begin{figure} 
\begin{picture}(110,140) \put(18,-10.5){\includegraphics[width=0.40\textwidth]{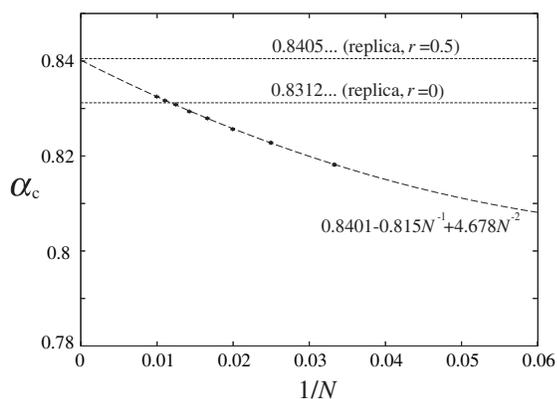}} 
\end{picture} 
\caption{Results of $L_1$-norm reconstruction experiment. We set 
$\rho=0.5$ and $r=0.5$. For each $N$ (=dimension of the original information) we 
took the average over $10^5$ samples. The broken line indicates the scaling by 
quadratic function regression. For $N \rightarrow \infty$, we clearly see the increase 
in the reconstruction limit from the case of no correlation ($\alpha_c=0.8312...$ for 
$\rho=0.5$ \cite{KWT}), which is consistent with the result from replica analysis. 
The value at $N\rightarrow \infty$ from the extrapolation is $\alpha_c=0.84017(28)$, 
which is very close to the value from replica analysis.} 
\label{fig:figure3} 
\end{figure}

For verification of the results from replica analysis, we also conducted a 
numerical experiment of the $L_1$-norm reconstruction. 
We used the convex optimization package for MATLAB developed in 
\cite{GB1,GB2}. The results are shown in Fig. \ref{fig:figure3}. The 
dependence of $\alpha_c$ on the dimension of the original coefficient $N$ is shown. 
For comparison with replica analysis, we also performed scaling analysis with quadratic 
function regression, and estimated the value of $\alpha_c$ for $N \rightarrow 
\infty$ limit. The results give $\alpha_c=0.84017(28)$ for $N \rightarrow \infty$, 
which clearly indicates the increase in the value of $\alpha_c$ (or degradation of the 
reconstruction performance) as expected. The reconstruction limit estimated by 
extrapolation is very close
to the one from replica analysis, which enforces the validity 
of the result from replica analysis. 



\section{Conclusions and Discussion}
We investigated the performance of the $L_p$-norm (especially $L_1$-norm) 
reconstruction with a correlated compression matrix by replica analysis. We 
obtained the expression of the cost function for the Kronecker-type compression matrix 
for general correlation matrices $\Rt$ and $\Rr$.

The noteworthy issues in the results are summarized as follows. First, the cost function 
does not depend on the correlation matrix $\Rr$ describing the correlation between 
observation vectors, which indicates that the observation 
procedure is not significant in CS. This can also be understood from the fact 
that we can eliminate the correlation matrix $\Rr$ by redefinition (or rotation) of the 
random matrix $\bm \Xi$.

Second, we considered the correlated compression matrix, whereas the 
original coefficient is uncorrelated. However, our result can be reinterpreted as the case 
of a correlated original coefficient and uncorrelated expansion bases by 
redefinition (or rotation) of these quantities.

Finally, we found that the performance of the $L_1$-norm reconstruction is robust 
against the small correlation between adjacent expansion bases of the original 
coefficient. By incorporating the strong correlation, the reconstruction 
performance slightly falls. This is a quite natural result because the correlation between 
expansion bases implies the loss of the original information by redundant bases, which 
makes the reconstruction much more difficult. We should also keep in mind that 
we discussed the reconstruction limit only in the case that the correlation matrix is 
tridiagonal, and there is a possibility that another correlation matrix might highly 
degrade the performance (note that our analysis offers the result for general correlation), 
which will be for future work.


\section*{Acknowledgments}
This research is supported by a Grant-in-Aid Scientific Research on Priority 
Areas ``Deepening and Expansion of Statistical Mechanical Informatics (DEX-SMI)`` 
from MEXT, Japan No. 18079006. Y.K. is also supported by the JSPS Global COE 
program, ``Computationism as a Foundation for the Sciences''.


\begin{thebibliography}{1}

\bibitem{CRT2006} 
E. J. Cand\`{e}s, J. Romberg and T. Tao, ``Robust uncertainty principles: Exact signal 
reconstruction from highly incomplete frequency information,'' IEEE Trans. Inform. 
Theory, vol. 52, no. 2, pp. 489-509, Feb. 2006. 

\bibitem{D2006-1} 
D. L. Donoho, ``Compressed sensing,'' IEEE Trans. Inform. Theory, vol. 52, no. 4, pp. 
1289-1306, Apr. 2006. 

\bibitem{CT2006} 
E. J. Cand\`{e}s and T. Tao, ``Near optimal signal recovery from random projections: 
Universal encoding strategies?,'' IEEE Trans. Inform. Theory, vol. 52, no. 12, pp. 
5406-5425, Dec. 2006. 

\bibitem{T2002} 
T. Tanaka, ``A Statistical-Mechanical Approach to Large-System Analysis of CDMA 
Multiuser Detectors,'' IEEE Trans. Inform. Theory, vol. 48, no. 11, pp. 2888-2910, Nov. 
2002. 

\bibitem{TUK} 
K. Takeda, S. Uda and Y. Kabashima, ``Analysis of CDMA systems that are 
characterized by eigenvalue spectrum,'' Europhys. Lett., vol. 76, no. 6, pp. 1193-1199, 
Dec. 2006.

\bibitem{HTK} 
H. Hatabu, K. Takeda and Y. Kabashima, ``Statistical mechanical analysis of the 
Kronecker channel model for MIMO wireless communication,'' Phys. Rev. E, vol. 80, 
061124, Dec. 2009. 

\bibitem{KWT} 
Y. Kabashima, T. Wadayama and T. Tanaka, ``A typical reconstruction limit for 
compressed sensing based on $L_p$-norm minimization,'' J. Stat. Mech., L09003, Sep. 
2009. 

\bibitem{D2006-2} 
D. L. Donoho, ``High-Dimensional Centrally Symmetric Polytopes with Neighborliness 
Proportional to Dimension,'' Discrete Comput. Geom., vol. 35, no. 4, pp. 617-652, 2006. 

\bibitem{DT2009} 
D. L. Donoho and J. Tanner, ``Counting faces of randomly projector polytopes when the 
projection radically lowers dimension,'' J. Amer. Math. Soc., vol. 22, no. 1, pp. 1-53, Jan. 
2009. 

\bibitem{RFG}
S. Rangan, A. K. Fletcher and V. K. Goyal, ``Asymptotic Analysis of MAP
	Estimation via the Replica Method and Compressed Sensing,''
 NIPS 2009, pp. 1545-1553.

\bibitem{GB1}
M. Grant and S. Boyd, { CVX: Matlab software for disciplined convex programming} 
(web page and software) 2009. http://stanford.edu/\~{ }boyd/cvx 

\bibitem{GB2} 
M. Grant and S. Boyd, { Graph implementations for nonsmooth convex programs}, 
Recent Advances in Learning and Control (a tribute to M. Vidyasagar), V. Blondel, S. 
Boyd, and H. Kimura, editors, pp. 95-110, Lecture Notes in Control and Information 
Sciences, Springer 2008. http://stanford.edu/~boyd/graph\b{ }dcp.html

\end{thebibliography}
\end{document}